\documentclass[%
 reprint,
 amsmath,amssymb,
 aps,
]{revtex4-1}

\usepackage{mathtools}
\usepackage{multirow}
\usepackage[export]{adjustbox}
\usepackage{graphicx}
\usepackage{dcolumn}
\usepackage{bm}
\usepackage{footmisc}
\usepackage[english]{babel}
\usepackage{natbib,hyperref}
\usepackage[normalem]{ulem}
\usepackage{comment}
\usepackage{xcolor}
\usepackage{subcaption} 
\usepackage{booktabs} 
\usepackage{algpseudocode}

\newcommand{\defeq}{\overset{\text{\tiny def}}{=}}

\DeclareMathOperator*{\argmin}{arg\,min}

\begin{document}

\preprint{APS/123-QED}
\graphicspath{ {./images/} }

\title{A Stochastic Compound Failure Model for Testing Resilience of Autonomous Fixed-Wing Aircraft I: Formulation and Simulation}
\author{Thelonious Cooper and Sai Ravela\\
Earth Signals and Systems Group\\
Massachusetts Institute of Technology\\
77 Massachusetts Avenue, Cambridge, MA 02139}

\begin{abstract}
    This paper presents a Markov chain model to dynamically emulate the effects of adverse (failure) flight conditions on fixed-wing, autonomous aircraft system actuators. It implements a PX4 Autopilot flight stack module that perturbs the attitude control inputs to the plane's actuator mixer.  We apply this approach in simulation on a fixed-wing autonomous aircraft to test the controller response to stochastic compound failures on a range of turning radii. Statistical measures of the differences between target and simulated flight paths demonstrate that a well-tuned PID controller remains competitive with adaptive control in a cascading, compound, transient failure regime.
\end{abstract}

\maketitle

\section{Introduction}

Despite their inadequacies in adverse conditions, low-cost fixed-wing Uncrewed/autonomous Aircraft Systems (UAS) are standard in environmental research. Where satellite imagery is too infrequent, strapping sensors to an inexpensive UAS and flying them is an attractive means for collecting data. In these missions, aircraft damage and payload losses are substantial when parts fail, even when the UAS is repairable. For missions such as flying into a volcanic plume or shallow cumuli~\cite{ravela13,Ravela2018}, errant ash or soot jamming a servo could mean data loss and thousands of dollars. 

In such flights, numerous things can go wrong. Electrical or communication failures induced by the environment, low-quality servos and cabling, poor thermal management, poorly strapped payload, and even mismatched wing surfaces are all culprits. In many situations, the failures are not immediately catastrophic but instead appear as incipient transient disturbances that cascade into compound failures, leading to total failure and permanent loss as too many things go wrong. Cascading failures often turn initial pilot or operator confusion or ignorance into panic, contributing to a crash.  

While redundant controls and sensors could render the aircraft flyable with degraded performance when a few parts fail, for autonomous or aided flight, improving autopilot resilience, handling, and recovering from developing transient in-flight failures would tremendously advance operating low-cost UAS in adverse environments.  However, this is easier said than done. Hierarchical PID controllers, at the heart of most stock modern hobby/environmental research uncrewed/autonomous aircraft autopilots~\cite{px4}, are generally well-tuned. Still, conventional wisdom is that finely tuned autopilots are not robust to ``incessant hiccups." For example, we do not expect stock hierarchical PID controllers to adapt quickly to changing vehicle dynamics or failing actuators; some adaptive control appears to be needed.

One approach is carefully modeling aircraft dynamics under various failure conditions and detecting them in flight to trigger an appropriate control law. This might involve model predictive control with online model parameter tuning and additional PID coupling to handle model imperfections. However, such approaches are likely to be very expensive. Adaptive control must be amenable to low-cost hardware in the envisioned use cases. In this context, exciting new developments in Retrospective Cost Adaptive Control (RCAC)~\cite{goel2020adaptive,goel2021a,goel2020a,lee2021adaptive} suggest tuning PID gains in a data-driven manner by optimizing a performance index on a retrospective performance variable. The RCAC solution recursively updates gains and is easy to implement on the typically low-cost hardware that stock autopilots, such as PX4~\cite{px4,px4reference}, used in this research, execute on. 
\begin{figure*}[htb!]
    \includegraphics[height=2.1cm]{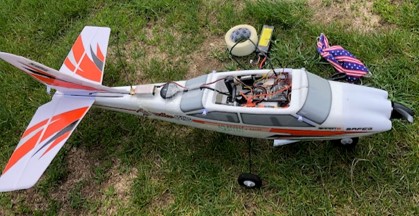}
    \includegraphics[height=2.1cm]{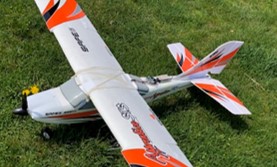}
    \includegraphics[height=2.1cm]{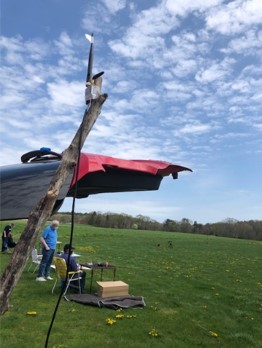}
    \includegraphics[height=2.1cm]{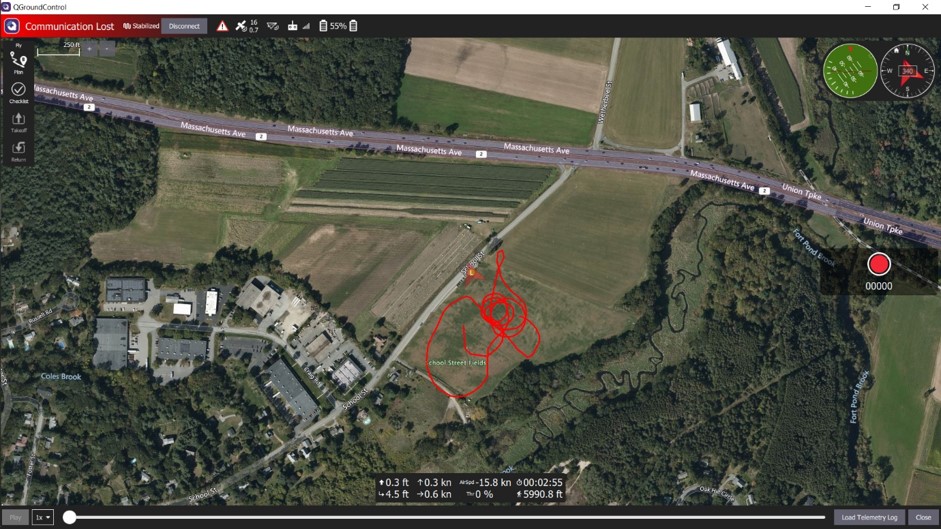}
    \includegraphics[height=2.1cm]{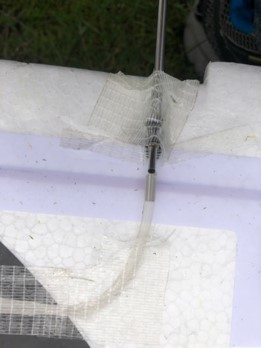}
    \includegraphics[height=2.1cm]{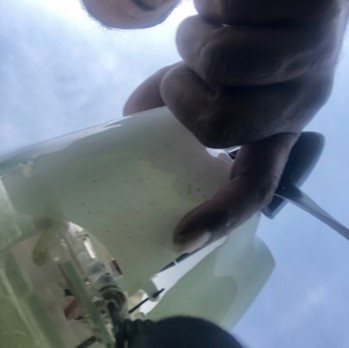}
    \caption{The Earth Signals and Systems group flies multiple aircraft for environmental applications, testing new control laws. Here, we show snapshots of a flight test with RCAC~\cite{lee2021adaptive}. From left to right, the Pixhawk autopilot mount, the aircraft, ground station, flight path, pitot tube assembly, and broken nose wheel. Such conditions as taped pitot tubes and broken assemblies are common in hobby flying.}
    \label{essg-flight}
\end{figure*}

Previously, we implemented RCAC for quadrotor~\cite{goel2020adaptive,goel2021a,spencer2021a} and fixed-wing aircraft~\cite{lee2021adaptive} in the PX4 environment (PX4 v1.10)~\cite{px4,px4reference} with Holybro Pixhawk (FMUv5)~\cite{px4HolybroPixhawk} hardware, and have tested them in software-in-the-loop, hardware-in-the-loop, and assisted flight configurations (see Figure~\ref{essg-flight}). In these tests, we flew the aircraft in {\em clean} and {\em adverse} (simulated failure) conditions, where cutting an actuator simulates failure. These configurations tested single-step failure injection. Based on our flight testing experience with fixed-wing aircraft, we posit that in many real-world cases, cascading, compounding intermittent failures that, left unresolved, become total failures are common. 

The current storyline starts with a well-configured aircraft and a well-tuned autopilot. That is, we are generally satisfied with the autopilot in clean conditions. We augment this autopilot with an adaptive controller (RCAC) so that the system shows some skill at adapting and maintaining performance in a demanding failure regime that is essential to test before use. In this regime,  transient actuator failures emerge, typically milli-seconds to seconds long. 

Despite existing solutions for simulating sensor failures, systematic approaches to testing cascading and compounding actuator failures are not readily available for PX4-like stock autopilots in small UAS settings. Here, we develop a stochastic model wherein a stationary Markov chain simulates actuator failure. It transitions the aircraft from a safe state (the clean configuration) to various degrees of simulated failure conditions, locking and releasing a combination of actuators at any given time. In this first attempt, the Markov transition matrix only loosely corresponds to flight observations from extensive prior flight experience (for example, see Figure~\ref{essg-flight}). It uses a Hardware-in-the-Loop (HITL) environment to test simulated failures and trajectories for fixed-wing aircraft. 

Our central hypothesis tests whether RCAC significantly improves resilience over well-tuned stock autopilots in the transient, cascading, and compound failure regimes. Our results indicate that with specific hyper-parameter tuning, RCAC deftly handles step changes to flight configurations. Previous work~\cite{lee2021adaptive,spencer2021a} shows that mixing its control inputs with the stock PX4 autopilot~\cite{px4} has a particular advantage for compensating poorly calibrated PID parameters. However, we find no specific benefit in the transient failure regime at typical timescales of interest. The adaptations tightly tune the aircraft to specific regimes, making transitions in a cascading failure regime difficult. It is possible that the issue could be addressed by dynamic RCAC hyper-parameter tuning. However, no clear strategies have emerged as of this writing. Adapting to cascading failures arriving at different rates for different lengths induces a performance variance that degrades overall resilience. 

The remainder of this paper is as follows. Section~\ref{sec:rw} describes related work. Section~\ref{sec:controller} describes the stock PID (PX4 v1.10)~\cite{px4} controller and RCAC~\cite{lee2021adaptive}, and Section~\ref{sec:methods} describes flight-path generation, Markov failure model, and comparison measures. Section~\ref{sec:implement} discusses implementation details for the Holybro Pixhawk (FMUv5), and a Discussion and Conclusions follow.
    
\section{Related Work}
\label{sec:rw}

There several existing adaptive control techniques for controlling fixed-wing aircraft~\cite{nguyen2006a} such as sliding mode~\cite{yu2020a,fu2021a}, backstepping-based nonlinear controller~\cite{hirano2019a}, and MRAC-based control augmentation~\cite{xiong2020a}. Although approaches such as ensemble control~\cite{trautner2020informative} on embedded GPUs are advancing, computationally expensive detailed models are still needed. We use RCAC, which is low cost and uses the past measured data and control inputs to recursively estimate controller gains~\cite{rahman2017a,goel2021a,lee2021adaptive,spencer2021a}. Extension to digital PID control~\cite{kamaldar2019a} and muticopter autopilot~\cite{goel2021a,spencer2021a} are available with an implementation for the PX4 Pixhawk platform. In previous work~\cite{lee2021adaptive}, RCAC was used to compensate for singleton actuator faults in both calibrated and degraded PX4 conditions~\cite{lee2021adaptive}, which we refer the reader to as the primary reference for details. Here, we use this framework to test performance over a stream of the transient compound, cascaded actuator faults. 

Prochazka et al.~\cite{prochazka_development_2021} develop a HITL demonstrator for validating fault-tolerant control methods for a hybrid UAV, where they simulate actuator failure and subsequent reconfiguration by a fault-tolerant flight control law. Jha et al.~\cite{jha18} develop a fault injector for autonomous vehicles, including data, hardware, timing, and machine learning subsystem components. Wen et al.~\cite{wen19} describe a fault simulator for UAS that includes sensor and actuator failures with human-in-the-loop flying. Gong et al.~\cite{gong19} argue that prior data is limited and develop a fault injector for UAS process health monitoring. In contrast, this work produces a stochastic Markov model for singleton, compound, and cascading actuator failures. Although unimplemented here, flight data can train the pre-specified transition probabilities. 

\section{Controllers}
\label{sec:controller}
Simulation experiments test two flight controllers,  the first is the stock PX4 autopilot based on hierarchical PID~\cite{px4}, and the second is RCAC~\cite{lee2021adaptive}. This section briefly discusses each controller, and a GitHub page~\cite{mitonrgit} contains the implementation of our work in the PX4 environment.

\subsection{PX4 PID Controllers for Fixed Wing Aircraft}

    The PX4 autopilot~\cite{px4reference} receives position setpoints from a Ground Controller (e.g., QGroundControl, Mission Planner) and controls the fixed-wing aircraft with two nested-loop cascaded controllers. A position controller operates the outer loop for longitudinal and lateral motion control, and an attitude controller operates the inner loop. The position controller's longitudinal control section sets the thrust ($T_s$) and pitch ($\theta_s$) set points, and the lateral control section sets the roll ($\phi_s$) set point. The attitude controller cascades two controllers. The first controller sets pitch ($\dot{\theta}_s$) and roll ($\dot{\phi}_s$) rates proportionately to roll and pitch errors (computing the yaw rate is algebraic). The second controller employs a feedforward and Proportional-Integral (PI) control law to set the angular acceleration. Control allocation methods then set actuator deflections using the angular acceleration set points.  

    Thus, the position controller takes the True airspeed ($V_{tas,s}$), attitude, and position setpoints and the corresponding measurements of true airspeed ($V_{tas}$), attitude, and position to set the thrust $T_s$, roll $\phi_s$, and pitch $\theta_s$ demands. Its longitudinal control section employs a Total Energy Control System (TECS)~\cite{bruce1986a,faleiro1999a,lambregts2013a,argyle2016a}, using a different guidance law~\cite{park2004a} for lateral control.
    
    The attitude controller's proportion control law uses gains $k_\theta$ and $k_\phi$ to produce rate setpoints:
    \begin{eqnarray}
        \label{eq:pitchrate}\dot{\theta}_s &=& k_\theta (\theta_s - \theta_m),\\
        \label{eq:rollrate}\dot{\phi}_s &=& k_\phi (\phi_s - \phi_m).
    \end{eqnarray}
    The yaw rate setpoint ($\dot{\psi}_s$) uses the true airspeed $V_{tas}$ and attitude to set develop a coordinated turn:
    \begin{equation}
        \dot{\psi}_s = \frac{g\tan{\phi_s}\cos{\theta_s}}{V_{tas}},
    \end{equation}
    where $g$ is gravity. The PI control law generates angular acceleration setpoint $\alpha_s$ as:
    \begin{equation}
    \alpha_s = \frac{\bar{V}_{tas}}{V_{tas}}\; k_{\omega,\text{ff}} \;\omega_s + \left(\frac{\bar{V}_{ias}}{V_{ias}}\right)^2 G_{\omega,\text{PI}}(\mathbf{q}) (\omega_s-\omega_m),
    \label{eq:angularacc}
    \end{equation}
    where $\bar{V}_{tas}$ and $\bar{V}_{ias}$ are the trim true and indicated airspeeds, respectively, $k_{\omega,\text{ff}}$ is the feedforward proportional gain, and $\omega_s$ and $\omega_m$ are the desired and measured body-frame angular velocities, given by 
    \begin{equation}
        \omega_s = \begin{bmatrix}
            1 & 0 & sin \theta \\
            0 & cos \phi & sin\phi \; cos\theta\\
            0 & - sin \phi & cos\phi\; cos\theta
        \end{bmatrix} \begin{bmatrix}
            \dot{\phi}_s \\
            \dot{\theta}_s \\
            \dot {\psi}_s
        \end{bmatrix}.
    \end{equation}

    The PI control law is: 
    \begin{equation}
        G_{\omega,\text{PI}}(\mathbf{q}) = k_{\omega,P} + {k_{\omega,I} \over \mathbf{q}-1},
    \end{equation}
    where $\mathbf{q}$ is the forward-shift operator. The gains $k_\theta$, $k_\phi$, $k_{\omega,\text{ff}}$, $k_{\omega,P}$, and $k_{\omega,I}$ constitute 11 variables that must be tuned for flight.
    
    \subsection{Retrospective Cost Adaptive Control for Fixed-wing Aircraft}

    Retrospective Cost Adaptive Control~\cite{lee2021adaptive} is a technique that uses a retrospective cost function to update the gains of the PID controller recursively. In the present application, we modify Equations~\ref{eq:pitchrate} and~\ref{eq:rollrate} as:
\begin{eqnarray}
        \label{eq:rcacptch}\dot{\theta}_s &=& k_\theta (\theta_s - \theta_m)+u_\theta,\\
        \label{eq:rcacroll}\dot{\phi}_s &=& k_\phi (\phi_s - \phi_m)+u_\phi.
    \end{eqnarray}
We modify Equation~\ref{eq:angularacc} as : 
 \begin{equation}
    \label{eq:rcacacc}\alpha_s = \frac{\bar{V}_{tas}}{V_{tas}}\; k_{\omega,\text{ff}} \;\omega_s + \left(\frac{\bar{V}_{ias}}{V_{ias}}\right)^2 G_{\omega,\text{PI}}(\mathbf{q}) (\omega_s-\omega_m) +u_{\omega,PI}.
    \end{equation}

    The control input terms $u_{\theta}$, $u_{\phi}$ and $u_{\omega,PI}$ are provided by the RCAC controller. 
\subsection{Adaptive Control Law}
    To understand how RCAC works, consider the example of a SISO PID controller with a feedforward term
    \begin{eqnarray}
        u_k &=& K_{p,k} e_{k-1} + K_{i,k} \nu_{k-1}\nonumber\\
        &&+ K_{d,k} (e_{k-1} - e_{k-2}) + K_{ff,k} r_k,
    \end{eqnarray}
    where $k\ge 0 $, $e_k$ is the error variable, $r_k$ is the feed-forward variable, and 
    \begin{eqnarray}
        \nu_k &\defeq & \sum_{n=0}^k g(e_n)\\
        &=& \nu_{k-1} + g(e_{k-1}).
    \end{eqnarray}

    A regression synthesizes the control law:
 \begin{equation}
        u_k = \gamma_k\rho_k,
    \end{equation} where \begin{equation}
        \gamma_k \defeq \begin{bmatrix}
            e_{k-1} & \nu_{k-1} & (e_{k-1}-e_{k-2}) & r_k
        \end{bmatrix}, \end{equation}and \begin{equation}\rho_k \defeq \begin{bmatrix}
        K_{p,k} &K_{i,k}& K_{d,k}& K_{ff,k}    
        \end{bmatrix}^T.
    \end{equation}

    To determine the controller gains $\rho_k$, consider a retrospective performance variable and error model~\cite{lee2021adaptive} \begin{equation}
        \hat{e}_k(\rho) \defeq e_k + \sigma (\gamma_{k-1}\rho - u_{k-1}),
    \end{equation}
    and the retrospective cost function $J_k$ \begin{eqnarray}  
    J_k(\rho) &=& \sum_{n=0}^k \hat{e}_n(\rho)^T R_{e} \hat{e}_n(\rho)\nonumber\\ 
    && + (\gamma_k\rho)^T R_u (\gamma_k\rho) \nonumber\\
&& +(\rho - \rho_0)^T P_0^{-1} (\rho - \rho_0).
    \end{eqnarray} 

    RCAC seeks the minimizer for $J_k$, that is \begin{equation}
        \rho_{k+1} \defeq \argmin_\rho J_k(\rho).
    \end{equation}

    For $k\ge 0 $, $\rho_{k+1}$ is recursively calculated~\cite{islam2019a}
    \begin{eqnarray}
        \rho_{k+1} = \rho_k \nonumber \\ && - \sigma P_{k+1} \gamma_{k-1}^T R_e [e_k + \sigma(\gamma_{k-1}\rho_k - u_{k-1})]
    \nonumber\\
    &&- P_{k+1}\gamma^T_k R_u \gamma_k \rho_k,
\end{eqnarray} 
where 
\begin{equation}
P_{k+1} = P_k - P_k \Gamma^T_k (R_{ue}^{-1} + \Gamma_k P_k \Gamma_k^T)^{-1}\Gamma_k,
\end{equation} and \begin{equation}\Gamma_k \defeq \begin{bmatrix}
    \sigma\gamma_{k-1} & \rho_k 
\end{bmatrix}^T, \end{equation} and \begin{equation}
 R_{ue} \defeq
    \begin{bmatrix}
       R_e & 0 \\
       0 & R_u
    \end{bmatrix}.
\end{equation} 
The following equation provides the control input \begin{equation}
    u_{k+1} = \gamma_{k+1}\rho_{k+1}
\end{equation}

The SISO case extends to MIMO through various parameterizations~\cite{goel2020a}. Here, RCAC modifies the PX4 attitude controller as shown in Equations~\ref{eq:rcacptch} - \ref{eq:rcacacc}, where the control inputs $u_\theta$, $u_\phi$, and $u_{\omega,PI}$ are generated for every discrete time step. We refer the reader~\cite{lee2021adaptive} for additional details of RCAC. The following sections discuss details of applying the stock PX4 and RCAC to flight tests.

\section{Methods}
\label{sec:methods}
Testing the stock and adaptive regimes requires three additional components. They include space-filling curves at different scales to express flight plans, the stochastic compound failure model, and a measure for comparing the desired and executed flight paths. These are discussed in this section.

\subsection{Hilbert Curves as Flight Test Patterns}

        A Hilbert curve~\cite{hilbert1891} is a continuous fractal space-filling curve, typically constructed as a limit of piecewise lines. The Hausdorff dimension of Hilbert curves is two, and its graph is a compact set homeomorphic to the closed unit interval, with Hausdorff dimension 2. The length of the $n^{th}$ curve is $2^n - {1 \over 2^n}$, i.e., the size grows exponentially with $n$, but each curve is in the unit square~\cite{wikipediaHilbertCurve}. The following pseudo code~\cite{mathworksHilbertCurve} is an easy example of Hilbert curves: 

        \begin{algorithmic}
             \State $[x,y] \gets Hilbert(n)$
             \If{$n\leq 0$}
             \State $x\gets 0$
             \State $y\gets 0$
             \Else
                \State $[xx,yy] \gets Hilbert(n-1)$
                 \State $x \gets \frac{1}{2}*[-\frac{1}{2}+yy \;\;-\frac{1}{2}+xx \;\;\frac{1}{2}+xx \;\;\frac{1}{2}-yy] $
\State  $y\leftarrow \frac{1}{2}*[-\frac{1}{2}+xx  \;\;\frac{1}{2}+yy \;\;\frac{1}{2}+yy \;\;-\frac{1}{2}-xx]$
  \EndIf
  \end{algorithmic}
        
\begin{figure*}[htb!]
    \centering
    \includegraphics[scale = 0.4]{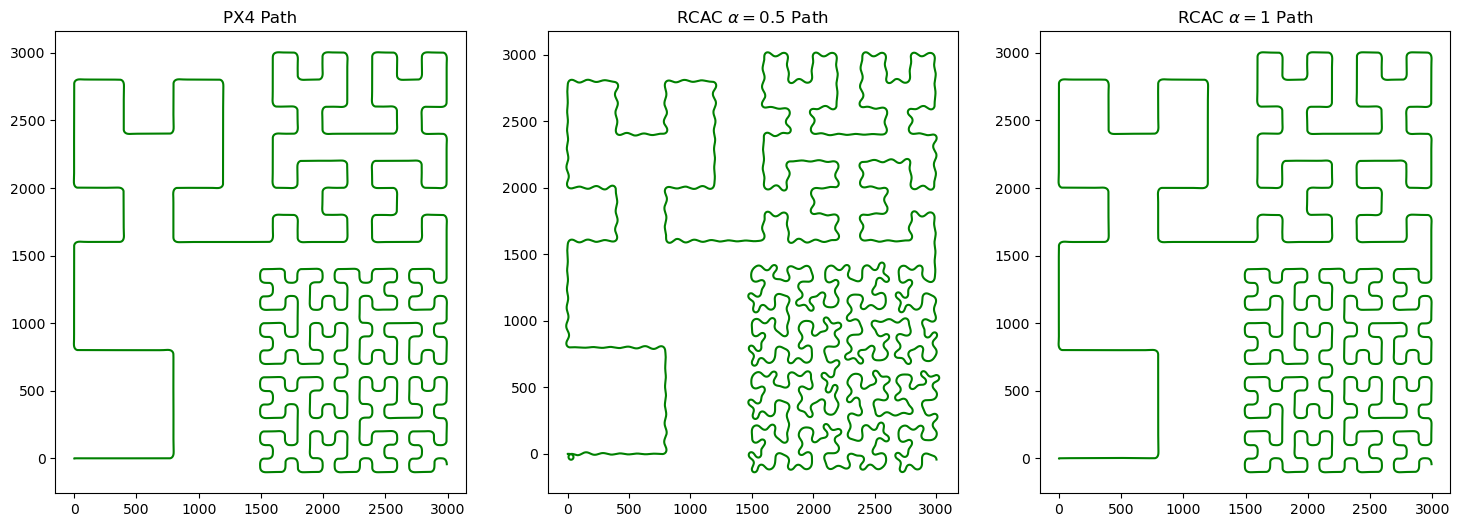}
    
    \caption{Examples of Nominal Flight Paths for PX4 (left) and RCAC (right).}
    \label{fig:hilbNominal}
    \end{figure*}

\begin{figure*}[htb!]
    
    \centering
    \includegraphics[scale = 0.4]{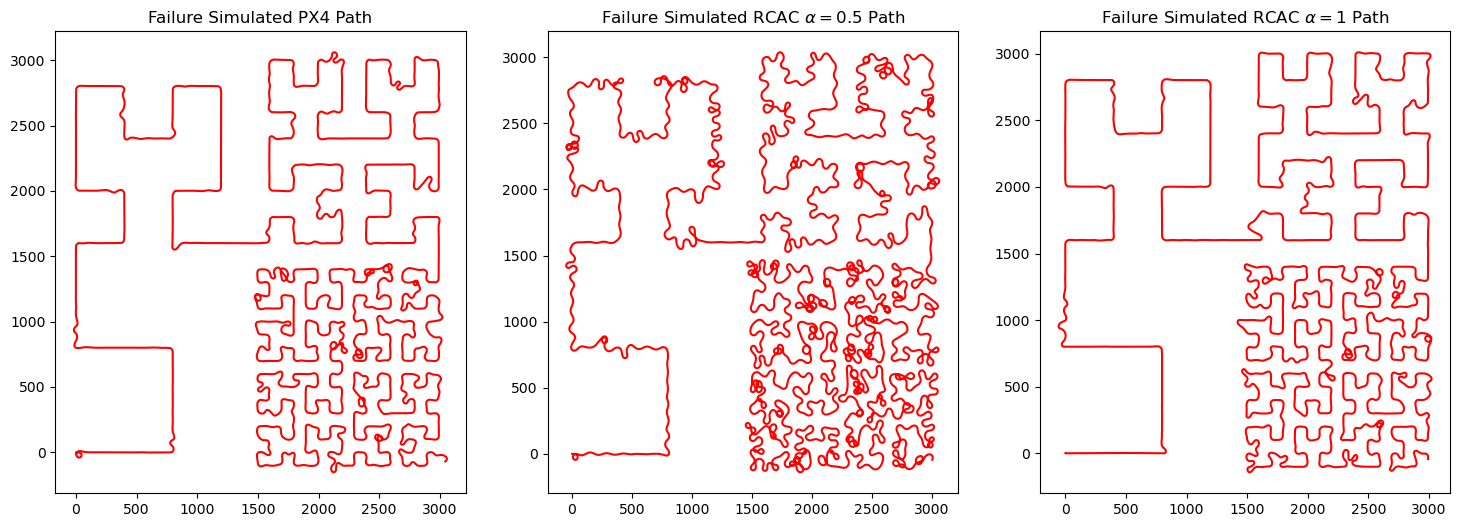}
    
    \caption{Examples of Failure Simulated Flight Paths for PX4 (left) and RCAC (right).}

    \label{fig:hilbPerturbed}
    \end{figure*}

    We designed a flight plan to test various turning frequencies based on Hilbert paths of varying order. We implemented the generation of these curves and used curves of orders 1,2,3 and 4 for each quadrant, respectively, see Figures~\ref{fig:hilbNominal} and ~\ref{fig:hilbPerturbed}.
    
    \begin{figure*}[htb!]
    \centering
    \includegraphics[scale=0.5]{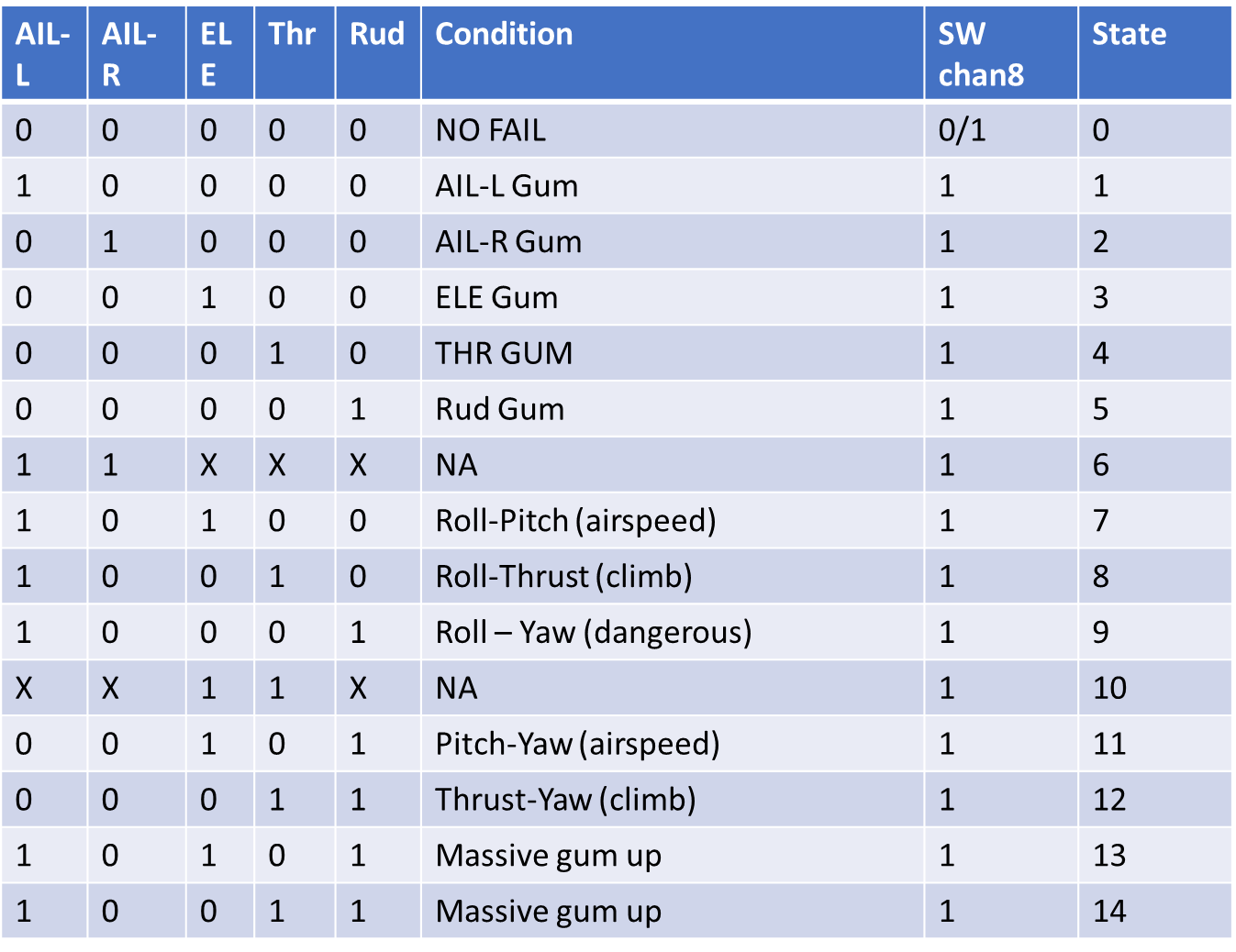}
    \caption{Failure states and the corresponding actuator failures. Each row depicts a Boolean condition indicating simulated failure (1) ``Gum" of particular actuators or combinations. An $X$ indicates that the row (failure) is never simulated (NA). The failure simulator proceeds if the handheld switch (SW on channel 8 in our setup) is on.  }
    \label{fig:failstate}
    \end{figure*}
    
    \begin{figure*}
    \centering
    \includegraphics[scale=0.4]{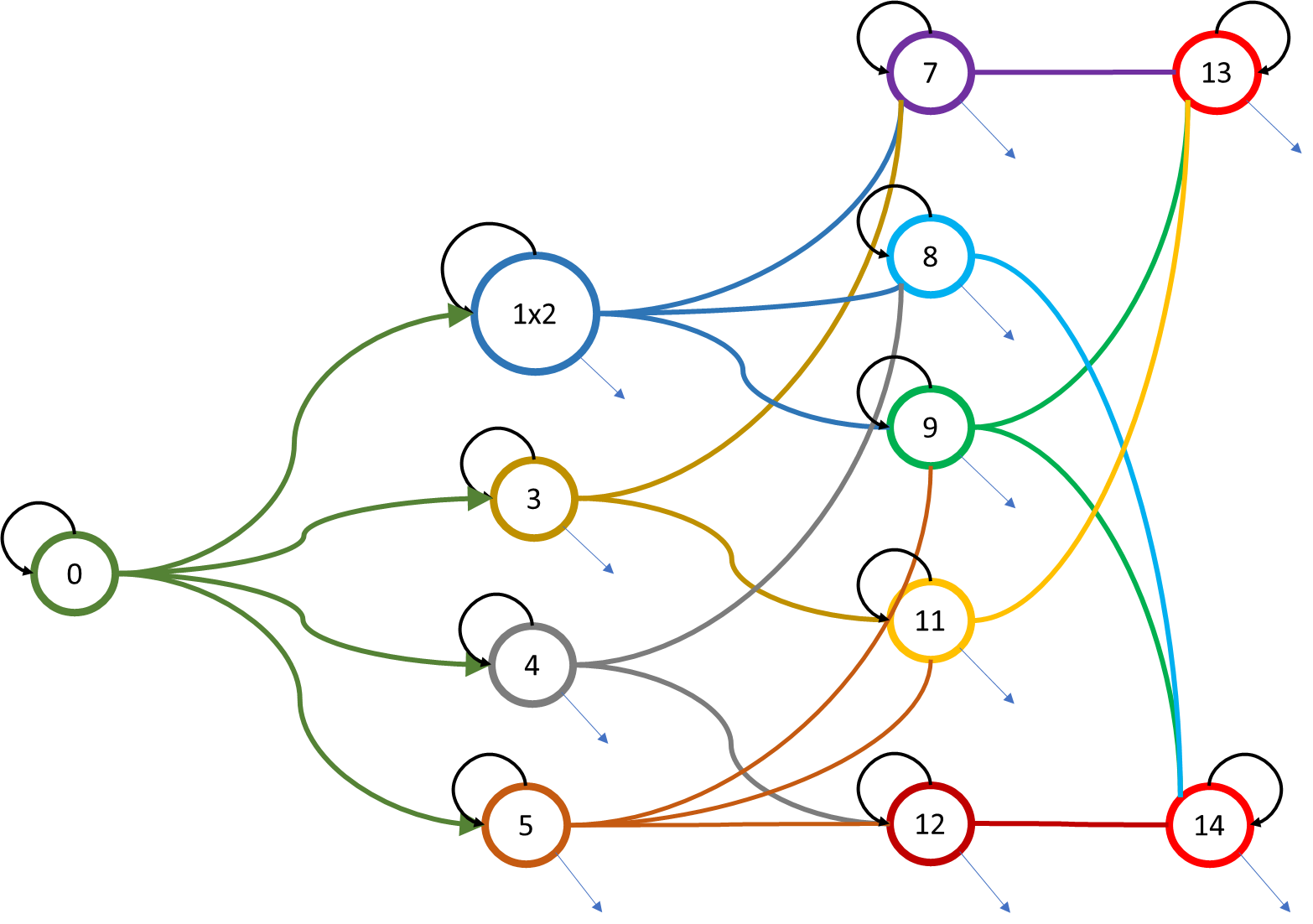}
    \caption{Markov process transition diagram, downward arrows represent edges to the ground state. The state numbers correspond to the table in Figure~\ref{fig:failstate}.} 
    \label{fig:markov}
    \end{figure*}
    
    \subsection{Simulating Compound Failures}

    The compound failure model tests multiple combinations of failures using a Markov chain. The ground state of {\it No Failure: 0} is continued indefinitely until a switch (S/W) on the ground handheld controller is activated to initiate failures. Turning the switch off reverts the aircraft to $State-0$ while continuing the flight. 
    
    Each transition from the ground state induces a single actuator failure. Although many forms of failure are admissible through our design, we only implemented the corresponding actuator getting stuck (``gummed") aileron left (AIL-L), right (AIL-R), an elevator (ELE), throttle (THR) or Rudder (Rud). Thus, the corresponding actuator is gummed up once transitioned to these states from the ground state. The aircraft may stay in this state, revert to State 0, or transition to compound failure states with one more failure. After that, transitions can continue through various combinations across states, each additional transition adding or removing a failure, in addition to staying in the same state or returning to the ground state. Certain compound states are never allowed. For example, State 10 in Figure~\ref{fig:failstate}, where both the elevator and throttle are stuck, will render the aircraft unrecoverable (depending on the throttle setting). Similarly, for field safety reasons, both ailerons are not failed (though one could still fly with the rudder in some cases). Thus, all the states are in some sense ``flyable" while still testing failures. 
    
    Our Markov chain simulations used a loop-back probability of $0.3$ and a ground-state transition probability of $0.4$, distributing the transition edges to valid additional failure states equally amongst the remaining probability. The transitions are discrete in time and triggered every second. Thus, the aircraft must stay in a state for at least one second with a $30\%$ chance of remaining there in the next second, $60\%$ possibility of transitioning to another failure, or a $40\%$ chance of returning to the ground state. In this way, not only is there persistence of a single failure, but various compound and intermittent failures are also triggered.

    Ensemble simulations of stochastic adverse flight conditions yield performance statistics richer than single-perturbation tests (e.g.,~\cite{lee2021adaptive}) for characterizing performance. In comparing different control schemes, each method can be subject to the same failure sequence to assess sample-wise and ensemble-wise differences. 
    
\subsection{Comparing Flight Paths Using Dynamic Time Warping}

    To compare flight paths, we resample the reference flight plan to the same length as the flown trajectory, saving it at nearly constant time intervals. Since our objective is to assess an ensemble of flight trajectories, the existence of relative timing differences can lead to the overall inflation of covariance~\cite{ravela2015stics}. The Kantorovich-Rubenstein or Wasserstein metric~\cite{kantorovich}, offers a natural approach to dealing with timing errors arriving at features (e.g., sharp turns). Here, we analyze the difference between the aircraft's performance in regular flight vs. facing a stochastic failure process using Dynamic Time Warping(DTW)~\cite{dtw01} between the flight paths. DTW was first proposed within the controls community in the 60s~\cite{bellman59} and has since enjoyed broad popularity in comparing discrete signals with index alignment errors. Note that DTW is not a metric, and in that, it differs from the Wasserstein distance. However, it is quick to compute, the values are usually close, and the reference signal (flight plan) is the same. A low DTW distance between the affected and unaffected flight paths represents that the autopilot could maintain course well despite the actuator failures. A high DTW distance implies the opposite. The nominal algorithm for DTW is as follows~\cite{wikipediaDynamicTime}:

    \begin{algorithmic}
        \State DTWfunc( $x[1\ldots m]$, $y[1\ldots n]$)
        \State $DTW[i,j] = \infty, \;\;\forall i=1\ldots m, j = 1\ldots n$
        \State $DTW[0,0] = 0;$
        \For {i=1:m}
        \For {j=1:n}
        \State $DTW[i,j] = Distance(x[i],y[j])+ \min (DTW [i-1,j], DTW[i,j-1], DTW[i-1,j-1])$
        \EndFor
        \EndFor
    \end{algorithmic} 
The function $Distance$ is a nominal point-wise Euclidean distance calculation or a metric in Hilbert space. The all-pair calculations incur quadratic costs, and incorporating window/taper functions simplifies it.

\section{Implementation}
\label{sec:implement}
    \begin{figure*}[htb!]
    
    \centering
    \includegraphics[scale = 0.5]{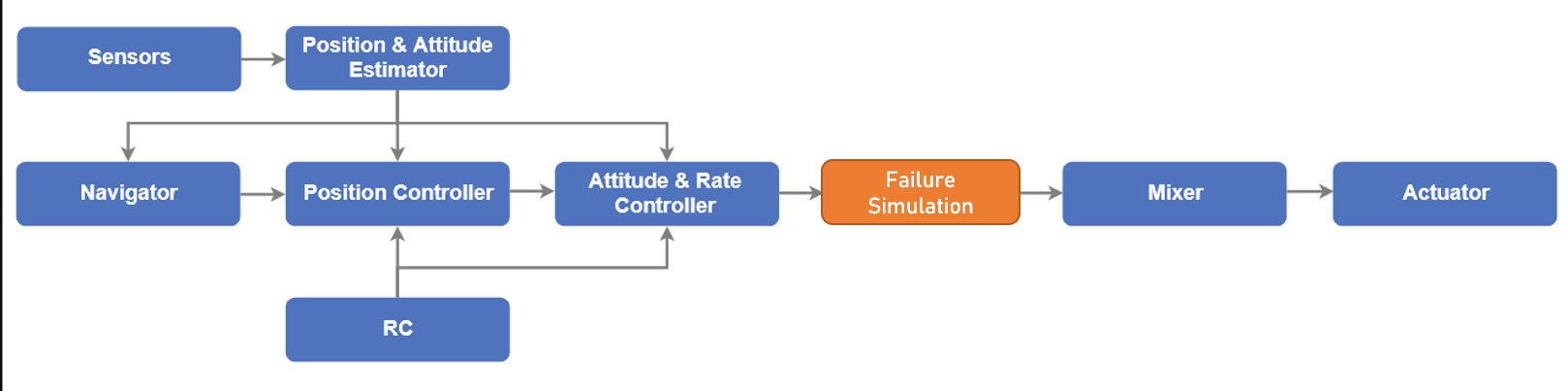}
    
    \caption{Modified PX4 architecture showing the failure injection module (in orange).}
    \end{figure*}
    
    To implement the control perturbation, we created a new instance of the actuator\_controls channel in the publish-subscribe px4 model. We then added a vehicle command that would tell the simulator or mixer to listen to the new instance of the actuator\_controls topic. Our module then listens to the original actuator\_controls topic and republishes it or sends a zero indicating a broken link. Besides creating our module, we had to modify the px4 architecture by altering the simulator module to listen for a command to switch control topic instances.
\begin{figure}[htb!]
    \includegraphics[width=2.5in]{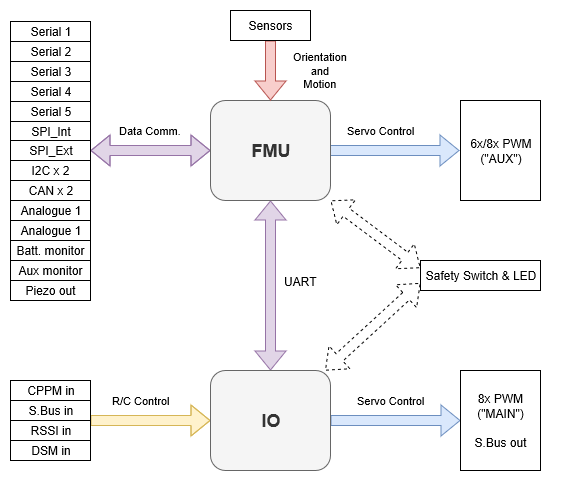}
    \caption{The PX4 implementation~\cite{px4reference} on PixHawk~\cite{px4HolybroPixhawk} splits the computation between two processors. Failure streams were re-implemented on FMU because the IO processor was weaker and could not sustain the computation.}
\end{figure}
    The Pixhawk PX4 (model) we tested with hardware in the loop testing uses two processors to perform flight control. The primary processor handles all of the mathematical computation, and a weaker coprocessor handles output operations and relays sensor information to the primary processor. We initially tried implementing the failure simulation on the IO coprocessor. However, this attempt failed because the additional processing of the failure simulation caused the coprocessor to break its real-time condition and crash. The IO coprocessor operates by a serial interface that blind writes into hard-coded memory sections which a critical section reads before being sent to actuators and other output devices.

    One challenge we encountered and could not solve fully was a race condition within the DDS data bus when stopping the failure simulation. Due to the real-time constraints of the system, after instructing the failure simulation process to stop, clock cycles are offered for a limited amount of time before deallocating its memory. This time is, unfortunately, faster than the update rate of the simulator, which waits for a signal to switch back to the original control signal instance only after attempting to read from the deallocated phony control signal buffer. That meant we could not stop the failure simulation process once it started. We could, however, command it not to modify the control inputs it parses and pass them along to the mixer unmodified, effectively disabling the purpose of the process.
\section{Experiments}
    

The PX4 autopilot is well-tuned for many hobby UAS. We calculated these gains for the aircraft under consideration~\ref{essg-flight} through extensive flight testing starting from readily available gains for similar aircraft. For RCAC, we performed several flight tests. We checked whether the parameter led to unstable flight in assisted mode with random flight-stick inputs for each controlled attitude variable. In many cases, the onset of instability occurred a considerable time after the stick perturbation. Therefore, we set nominal values, which may slow the gain adaptation and be sub-optimal. Table~\ref{tab:rcacpar} lists the settings we used. 

\begin{table}[htb!]
    \centering
    \begin{tabular}{|c|c|c|} 
\hline
PTCH  & P0 & 1 \\
\hline
PTCH & RU & 0.001 \\
\hline
PITCH RATE& P0& 1e-4 \\
\hline
PITCH RATE & RU &0.1\\
\hline
ROLL & P0& 1\\
\hline
ROLL  & RU &0.001\\
\hline
ROL RATE & P0 &1e-4\\
\hline
ROLL RATE & RU &0.1\\
\hline
YAW RATE & P0 &1e-4\\
\hline
YAW RATE &RU & 0.1\\
\hline
\end{tabular}
    \caption{Nominal RCAC (hyper) parameters. The mixing variable $\alpha$ is to $1$ (or $0.5$).}
    \label{tab:rcacpar}
\end{table}

For the standard PX4 PID-based autopilot, we ran $8$ trials of the stochastic adversarial actuator failure process. For an experimental Retrospective Cost Adaptive Control(RCAC) algorithm, we ran $11$ trials with the mixing hyperparameter $\alpha = 0.5$ and 12 trials with $\alpha = 1$. As a baseline, we also ran several clean flights of the PX4 PID controller. We saw they had an average self-similarity under the DTW metric of $133$ meters with a standard deviation of $47$ meters.   We then did the same for the RCAC $\alpha=0.5$ autopilot and saw an average self-similarity under the DTW metric of $172$ meters with a standard deviation of $46$ meters. And then again did the same for the RCAC $\alpha=1$ autopilot and saw that they had an average self-similarity under the DTW metric of $136$ meters with a standard deviation of $35$ meters. 

    The results of our experiments are summarized in the following tables
\begin{table}[htb!]
    \centering
    \begin{tabular}{|c|c|c|} 
\hline
Flight Configuration  & Mean & Standard Deviation \\
\hline
Px4 PID & 461 & 39 \\
\hline
RCAC $\alpha=0.5$ & 604 & 45 \\
\hline
RCAC $\alpha=1$ & 457 & 19\\
\hline
\end{tabular}
    \caption{Mean and standard dev of DTW distance to target flight path for unperturbed autopilot configurations}
    \label{tab:rcacpar}
\end{table}

\begin{table}[htb!]
    \centering
    \begin{tabular}{|c|c|c|} 
\hline
Flight Configuration  & Mean & Standard Deviation \\
\hline
Px4 PID & 595 & 154 \\
\hline
RCAC $\alpha=0.5$ & 1196 & 302 \\
\hline
RCAC $\alpha=1$ & 610 & 163\\
\hline
\end{tabular}
    \caption{Mean and standard dev of DTW distance to target flight path for failure-simulated autopilot configurations}
    \label{tab:rcacpar}
\end{table}
    \begin{figure*}[htb!]
    \centering
    \includegraphics[scale = 0.5]{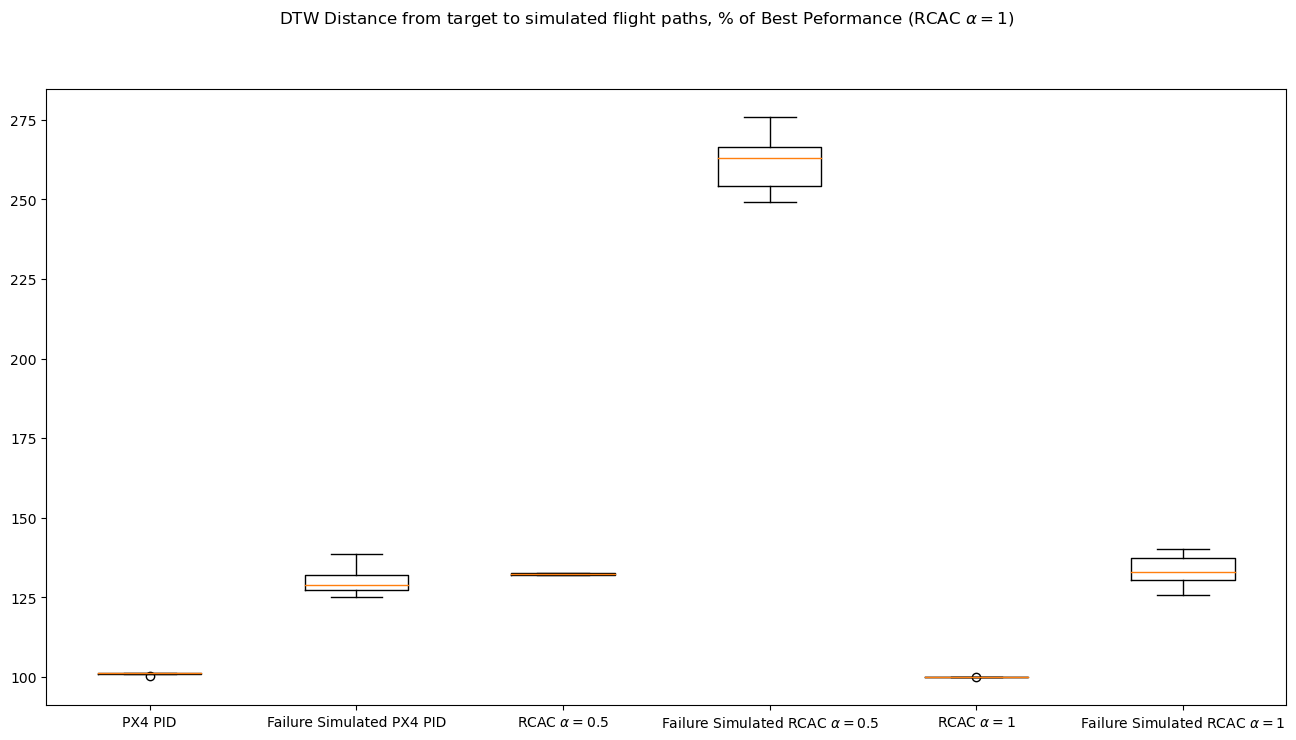}
    \caption{Box plot of DTW distances for nominal (clean) and failure simulated (adverse) flight conditions. }
    \label{fig:compareDTW}
    \end{figure*}
    \section{Conclusions and Future Work}
    This paper considers transient singleton, compound, and cascading actuator failure scenarios and tests the performance of a stock autopilot PX4 (v1.10) simulation for Holybro Pixhawk (FMUv5) hardware and fixed-wing aircraft. In actual flight experiments we have conducted elsewhere, multiple failures often cascade. For example, transient spikes in electrical load can trigger losses anywhere in the system. The PX4 autopilot, like others, is often tuned for the aircraft, and considerable attention is devoted to this effort before commencing flight missions. The case for extending RCAC-like adaptive control from tuning degraded gains to failure streams emerging in well-tuned autopilots is interesting and essential. The operational setting is of a tuned autopilot with the RCAC as ``safety" that would compensate for the transient failure stream. 
    
    In this paper, a Markov chain simulates failures with a stationary distribution that reaches all clean and adverse modes. The Markov chain is designed not to cause total catastrophic failure of the aircraft. Simulated Hilbert curves are flight paths for testing PX4 and PX4-RCAC, with mixing parameter $\alpha = 0.5$ and $\alpha = 1$. The PX4 flight parameters were determined for the model aircraft through flight testing and set in simulations. We selected RCAC hyerparameters that led to stable flight over long duration for the same model aircraft. DTW was used to compare resampled flight plans and simulated flight paths to account for timing-error artifacts between ensemble simulations. The ensemble simulations calculate the mean position error and the error variance. Under clean configurations, RCAC slightly outperforms the stock PX4 PID controller,  but under the hostile regime, it under performs. Additional, possibly dynamic, hyper-parameter tuning may be required to deal with intermittent failure streams. With the extra work, we will implement this in actual flight tests in the future. We also anticipate learning the Markov chain using flight logs gleaned from the PX4 community.
\section{Contributions and Acknowledgment}
    Corresponding Author: Thelonious Cooper ({\it theloni@mit.edu}). Thelonious Cooper wrote code and performed the experiments, Sai Ravela developed the transition tables, and Erina  Yamaguchi reported the original Hilbert path code. Cooper and Ravela drafted and edited the paper. We thank Prof. Ankit Goel, Jyonghoon Lee, Juan Augusto Paredes, John Spencer, and Prof. Dennis Bernstein for their support and collaboration. We are grateful for funding from ONR (N00014-19-1-2273) and the MIT Weather Extreme and CREWSNET Climate Grand Challenge projects.

\bibliography{refs,2110.11390}
\bibliographystyle{plain}
\end{document}